\documentclass{elsart}
\usepackage{graphicx,subfigure}
\usepackage{natbib}

\usepackage{amssymb}

\def\xmm{{\it XMM-Newton}}

\begin{document}

\begin{frontmatter}



\title{Merging clusters of galaxies observed with XMM-Newton}


\author[label1]{E. Belsole}
\author[label2]{J-L. Sauvageot}
\author[label3]{G.W. Pratt}
\author[label4]{and H. Bourdin}

\address[label1]{H.H. Wills Physics Laboratory - University of
  Bristol, Tyndall Avenue, Bristol BS8 1TL, UK} 
\address[label2]{Service d'Astrophysique, CE-Saclay, L'orme des
  Merisiers, 91191 Gif sur Yvette Cedex, France} 
\address[label3]{MPE Garching, Giessenbachstra{\ss}e, 85748 Garching,
  Germany} 
\address[label4]{Dipartimento di Fisica, Universit\`a degli studi di Roma Tor Vergata, Via della Ricerca Scientifica 1, 00133 Roma, Italy }  

\begin{abstract}

We present results from the XMM-Newton observations of our ongoing
program on merging 
clusters. To date three clusters have been observed, covering the
temporal sequence from early to late stage mergers: A1750, A2065 and
A3921. Using spatially-resolved spectroscopy of discrete regions,
hardness ratio and temperature maps, we show that all three clusters
display a complex temperature structure. In the case of A1750, a
double cluster, we argue that the observed temperature structure is
not only related to the ongoing merger but also to previous merger
events. A2065 seems an excellent example of a `compact
merger', i.e. when the centres of the two clusters have just started
to interact, producing a shock in the ICM. Using comparisons with 
numerical simulations and complementary
optical data, the highly complex temperature structure evident in A3921
is interpreted as an off-axis merger between two unequal mass
components. These results illustrate
the complex physics of merger events. The relaxation time can be
larger than the typical time between merger events, so that the
present day morphology of clusters depends not only on on-going
interaction but also on the more ancient formation history. 
\end{abstract}

\end{frontmatter}

\section{Introduction}\label{sec:intro}
As the largest assembled structures in the Universe, clusters of
galaxies are commonly thought to form by gradual accretion of matter
along filaments and by interaction and merging with previously formed
structures.  Hydrodynamical simulations (e.g. \citealt[]{rtk04} and
references therein) predict that merger events strongly affect the
physical characteristics of the intra-cluster medium (ICM). In
particular the temperature structure is thought to be an excellent
indicator of the cluster dynamical state and formation
history. Spectro-imaging observations with XMM-Newton and Chandra
allow the building of precise temperature maps, enabling deeper
investigation of the dynamical processes of cluster formation and the
effect of the mergers on the ICM
\citep[e.g.,][]{mark,neu,henry,kriv,vikh}.

With the aim of describing an evolutionary sequence of cluster
formation, we have selected a small sample of galaxy clusters showing
morphological evidence of ongoing merger activity. The sample was
selected, on the basis of previous X-ray observations, from nearby
clusters which entered the field of view of XMM-Newton.  Here we
summarise results for A1750 and A3921, presented in detail in
\citet[]{belsa,belsb}, and we describe preliminary, new results for
A2065.

\section{Observations and data analysis}

In Table \ref{tab:obs} we list the main physical characteristics of
the three clusters, together with basic observation information. 
The A1750 and A3921 observations were very little contaminated by soft
proton flares (details of the data preparation can be found in
\citealt{belsa,belsb}). Unfortunately, emission from
soft protons 
dominates the whole observation of A2065. This required an {\it ad-hoc}
treatment for these data, which allowed us to model the
background. A2065 will be re-observed with XMM-Newton, and
thus here we discuss only preliminary results. 

The background estimates were obtained using a blank-sky observation
consisting of several high-latitude pointings with sources removed
(\cite{lumbbkg}), and source and background events were corrected for
vignetting using the {\sc eviweight} task in the Science Analysis
System (SAS), enabling us to use the on-axis response matrices and
effective area.

\section{Results}

\begin{table}
\begin{minipage}{\columnwidth}
\begin{center}
\caption{Journal of observations.}\label{tab:obs}
\begin{tabular}{lccc}
\hline
Parameter       & Abell 1750 & Abell 2065  & Abell 3921 \\
\hline
$z$    & 0.086 &  0.072& 0.096 \\
RA (J2000) & $13^h30^m52^s$ & $15^h22^m42.6^s$ & $22^h49^m38.6^s$ \\ 
Dec(J2000) &$-01^o50'27''$  & $+27^o43''21''$ & $-64^o23'15''$ \\
Obs time (ks) & 30 & 23 & 30 \\
$N_{\rm H}$(10$^{20}$ cm$^{-2})$\footnote{From \citet{dlnh}} & 2.39 &
2.95 & 2.94 \\ 
Global T (keV) & 3.87 (2.84)\footnote{A1750C (A1750N)}& 5.4 & 5.0 \\
\hline
\end{tabular}
\end{center}
\end{minipage}
\end{table}

\begin{figure*}
\begin{minipage}{\columnwidth}
\centering
\hspace{-0.8cm}
\subfigure[] 
{
    \label{fig:fig1a}
    \includegraphics[scale=0.38,angle=0,keepaspectratio]{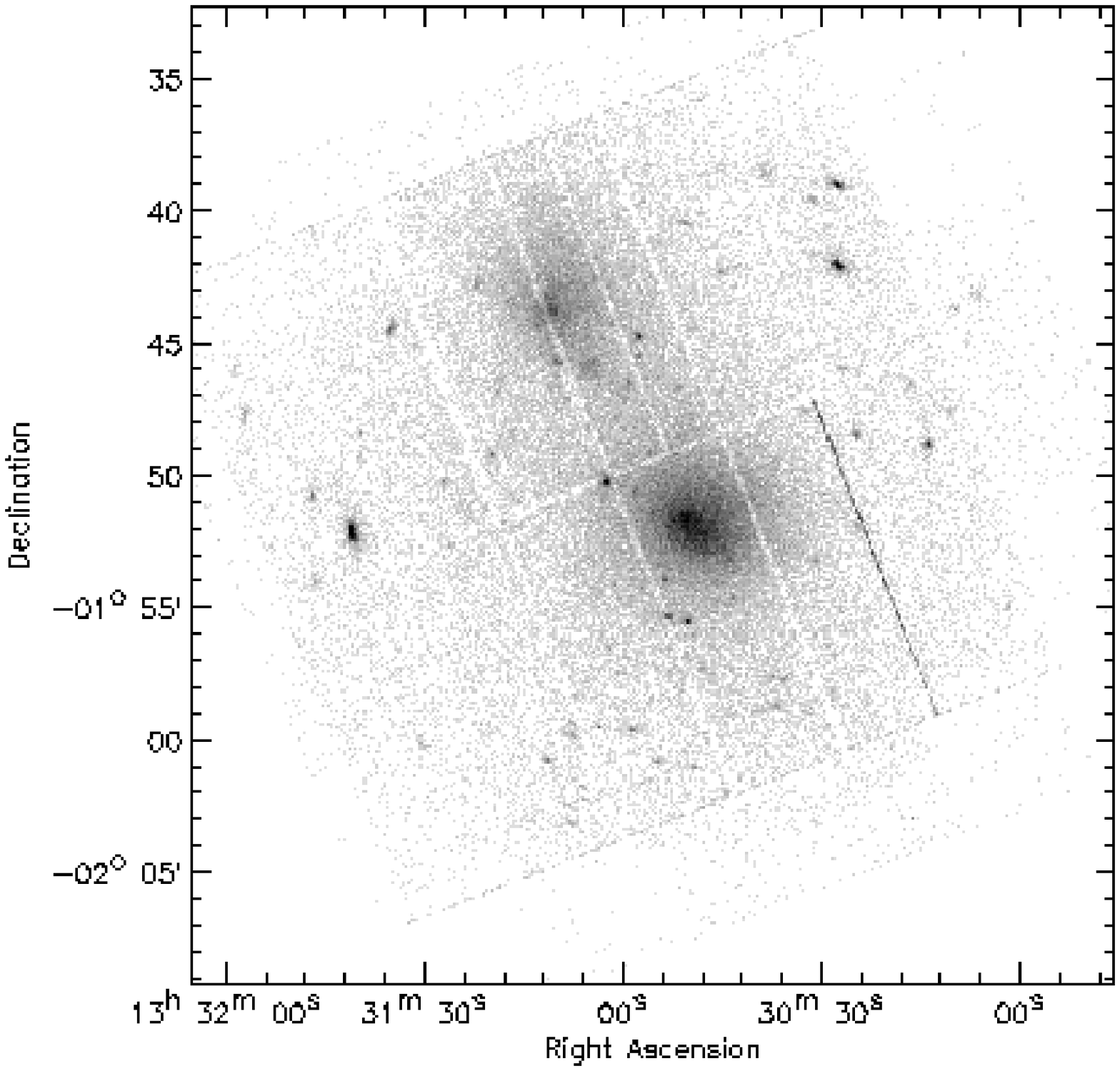}
}
\hspace{-0.4cm}
\subfigure[] 
{
    \label{fig:fig1b}
    \includegraphics[scale=0.38,angle=0,keepaspectratio]{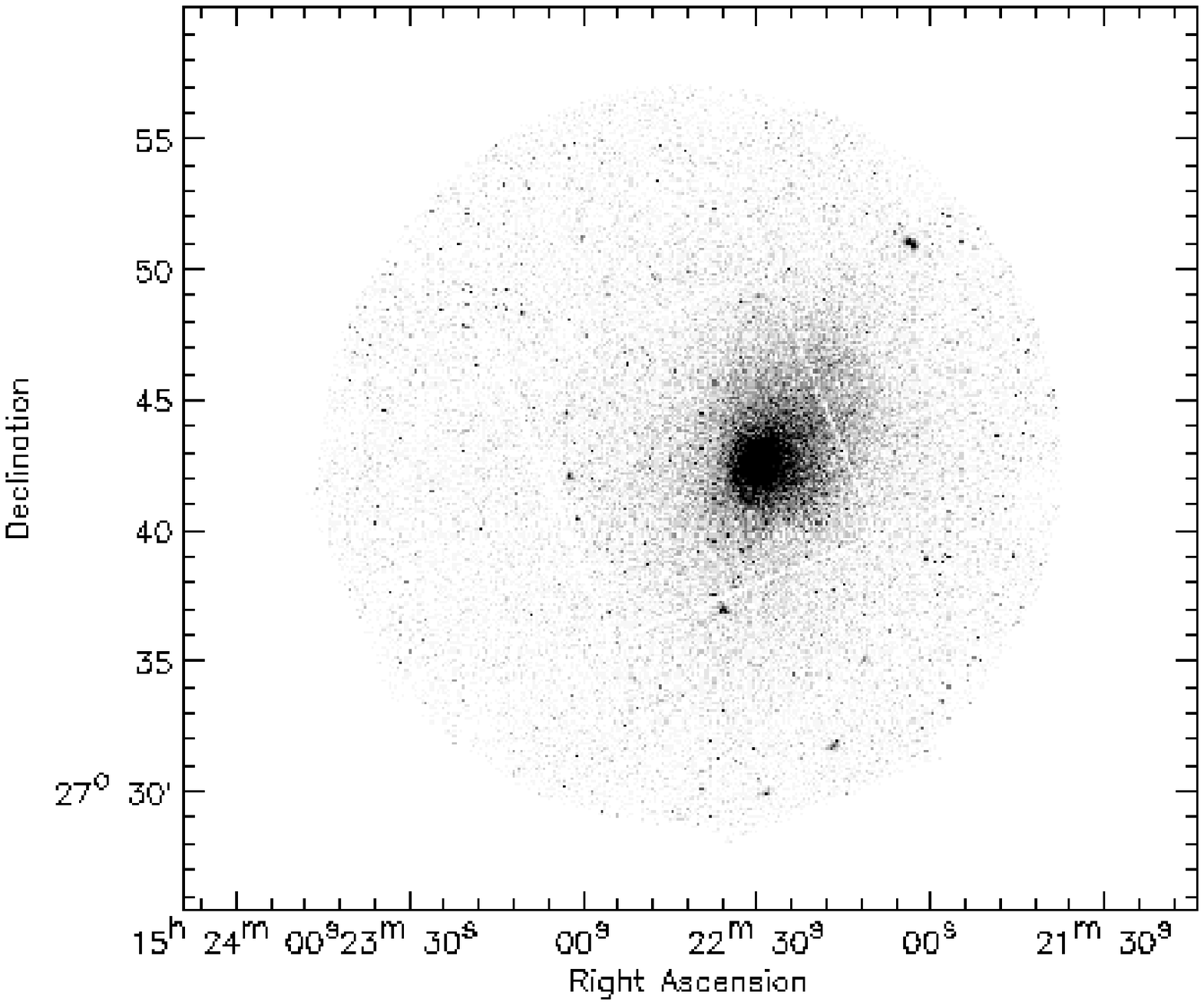}
}
\hspace{-0.5cm}
\subfigure[] 
{
    \label{fig:fig1c}
    \includegraphics[scale=0.38,angle=0,keepaspectratio]{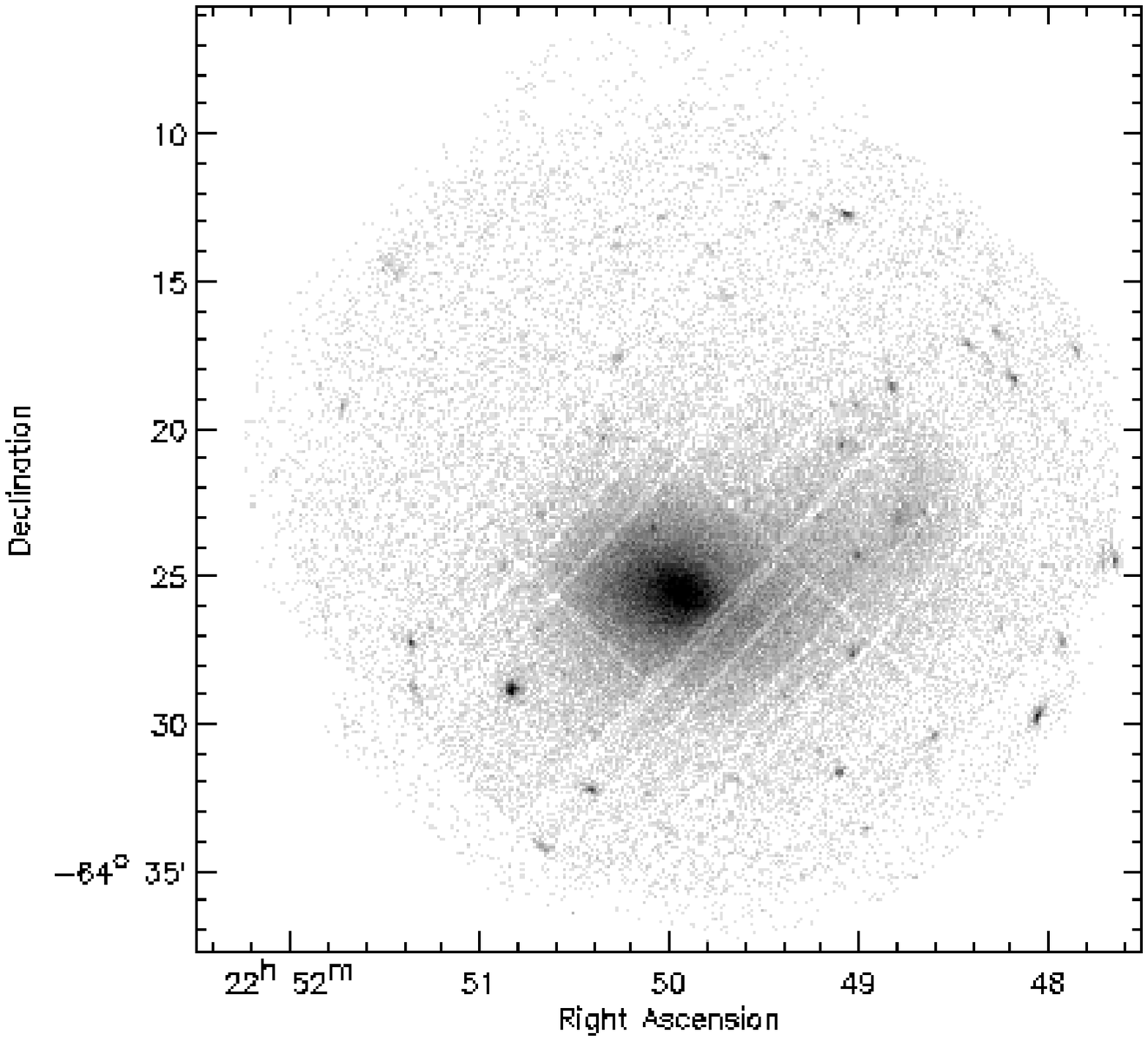}
}
\caption{{\footnotesize (a) \xmm\/EPIC  0.3-7.0 keV energy band count image of
    A1750; (b) MOS1+MOS2 0.1-1.4 keV energy band count  image of
    A2065, the band chosen to enhance cluster emission above the
    strong particle background; (c) \xmm\/EPIC 0.3-10 keV energy band count image of A3921. These 
    are raw, non-background subtracted images.}} 
\label{fig:rawima} 
\end{minipage}
\end{figure*}

We show in Figure \ref{fig:rawima} the images of A1750 and A3921 in
the 0.3-7.0 keV band; for A2065 a lower energy band [0.1-1.4 keV] is
shown to enhance cluster emission above the strong particle
background. 
We observe that the morphology of the clusters is very different,
reflecting already a difference in their dynamical state. 
A1750 is a prototype of a binary cluster, although the system appears
to have at least three components \citep{beers}. The main cluster,
A1750C, is at the centre of the field of view, and a sightly less
massive cluster (A1750N) lies to the north. A2065 looks rather
unperturbed on the outskirts, but the central regions display a clear
compression of the isophotes towards the south-east, and a sharp edge
to the east. Finally, the X-ray image of A3921 is dominated by the
main cluster at the centre of the field of view, and we observe an
elongation of diffuse emission to the west, where the X-ray emission
from two of the three brightest galaxies in the cluster can also be
seen. 

\subsection{Morphology}

One approach to evaluate any deviation from a dynamically relaxed
cluster is to assume a $\beta$-model \citep{cff} as a description of a
relaxed isothermal system. The large soft proton contamination of the
A2065 observation reduced the significance of any substructure when
this method was applied and thus we do not discuss it further.  

We thus generated low energy (0.3-2.0 keV, where the density
distribution is least temperature dependent) surface brightness maps
for A1750 and A3921, and fitted them with a bi-dimensional
$\beta$-model. The two clusters belonging to A1750 were taken into
account simultaneously. We found that A1750 does not show significant
residuals above a $\beta$-model, in the region between the two
clusters, but there is some excess emission in their cores.

On the other hand, A3921 shows large residuals to the west (other
than those in the centre of the main cluster), indicating that a
significant substructure is present and it is related to one or both
the brightest galaxies in that area. The two galaxies are at 7 arcmin (RA = $22^h48^m49^s$, Dec = $-64^\circ23^\prime10^{\prime\prime}$ (J2000)) to the north-west (BG2) and at 8 arcmin (RA = $22^h49^m04^s$, Dec = $-64^\circ20^\prime35^{\prime\prime}$ (J2000)) to the WNW (BG3) from the central brightest galaxy BG1 (RA = $22^h49^m58^s$; Dec=$-64^\circ25^{\prime}46^{\prime\prime}$ (J2000); see Belsole et al. 2005 for details). 
\begin{figure*}
\begin{minipage}{\columnwidth}
\centering
\subfigure[] 
{
    \label{fig:fig2a}
    \includegraphics[scale=0.53,angle=0,keepaspectratio]{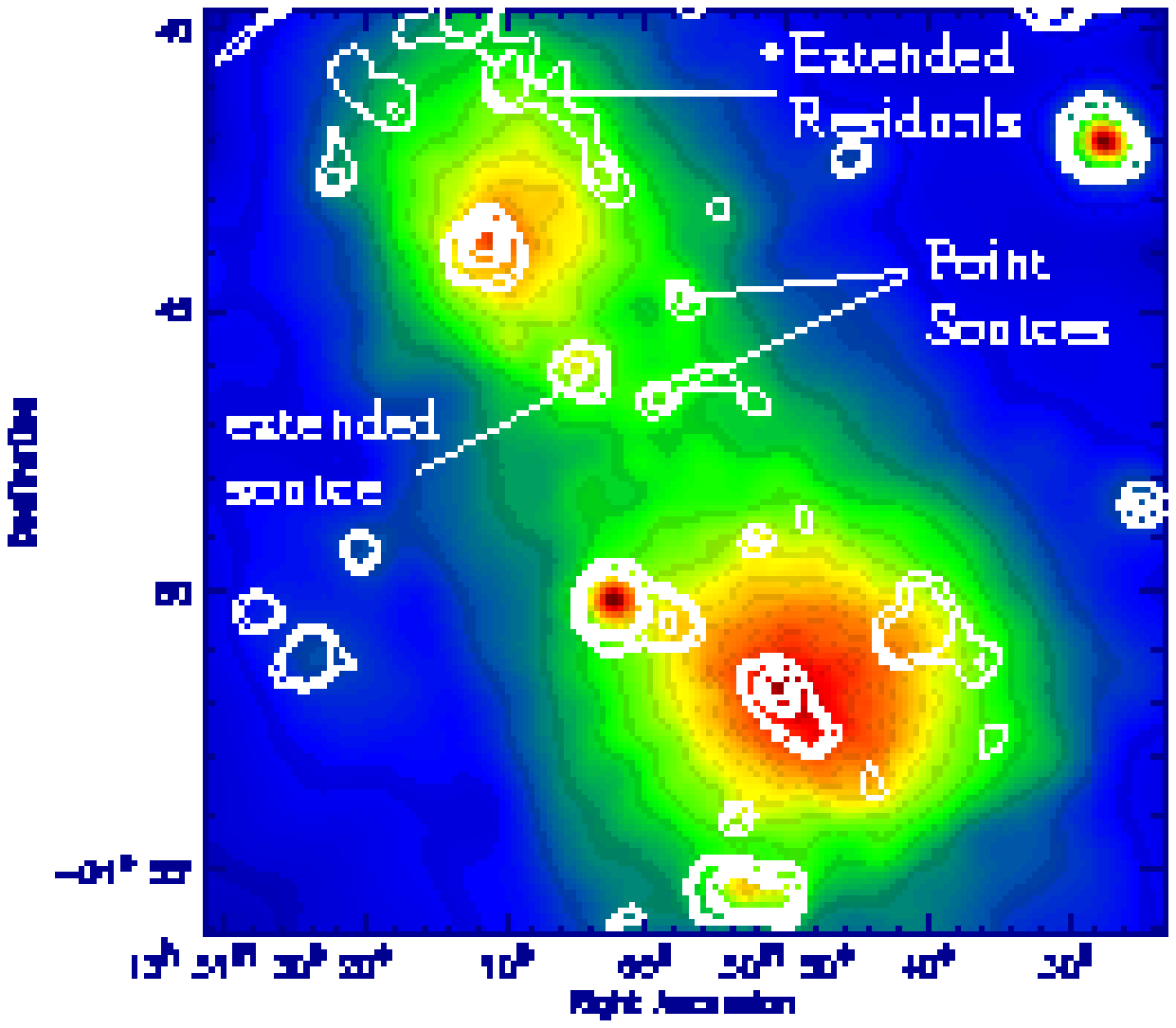}
}
\hspace{-0.5cm}
\subfigure[] 
{
    \label{fig:fig2b}
    \includegraphics[scale=0.37,angle=0,keepaspectratio]{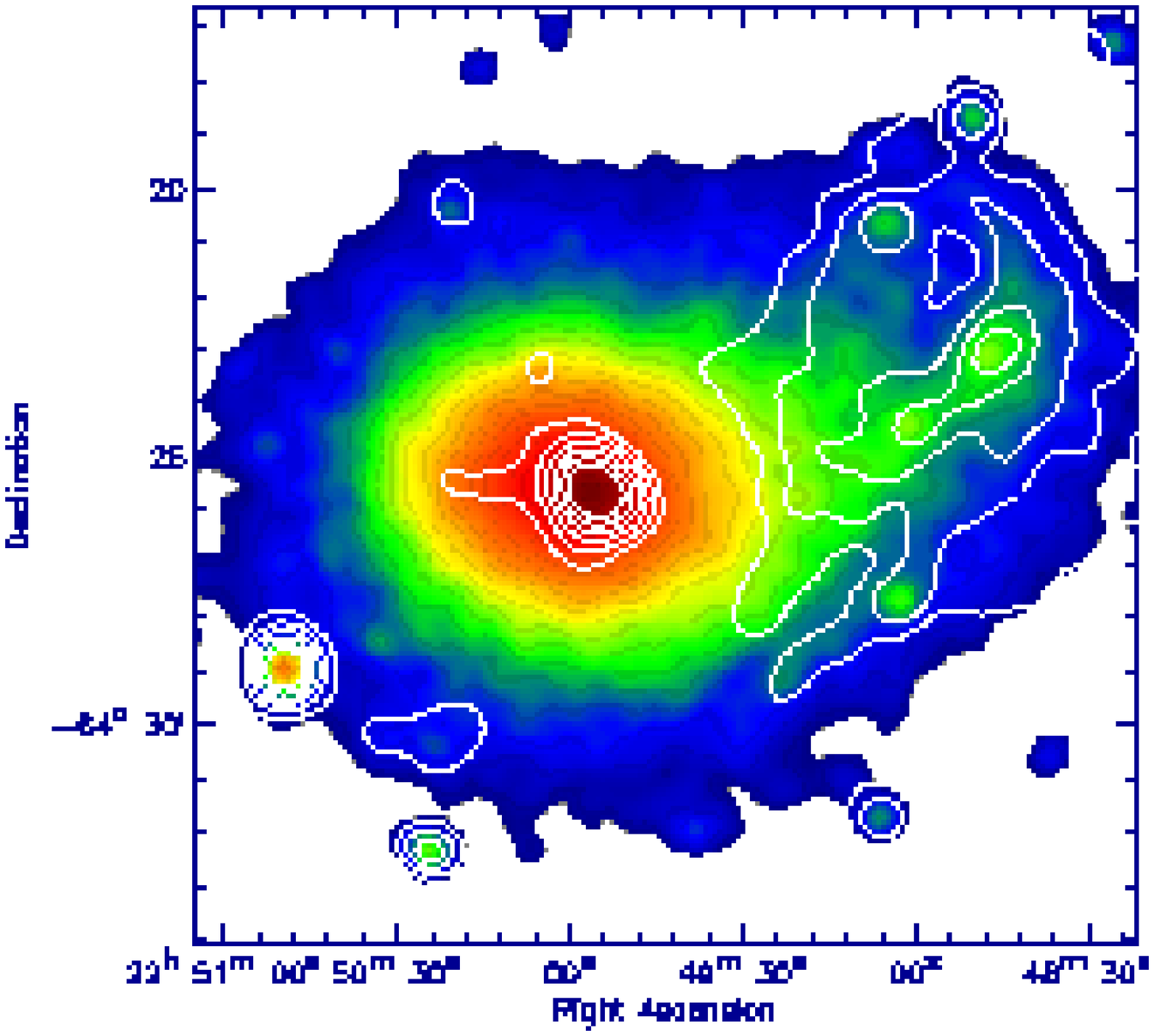}
}
\caption{{\footnotesize Surface brightness maps of A1750 (a) and
    A3921(b) in the soft energy band (0.3-2.0 keV). Contours are the
    residuals above a 2D $\beta$-model which was subtracted from the
    image.The first contour is at 1$\sigma$.}} 
\label{fig:betamod}  
\end{minipage}
\end{figure*}

\subsection{Temperature distribution}
\begin{figure*}
\begin{minipage}{\columnwidth}
\centering
\hspace{-0.9cm}
\subfigure[] 
{
    \label{fig:fig3a}
    \includegraphics[scale=0.38,angle=0,keepaspectratio]{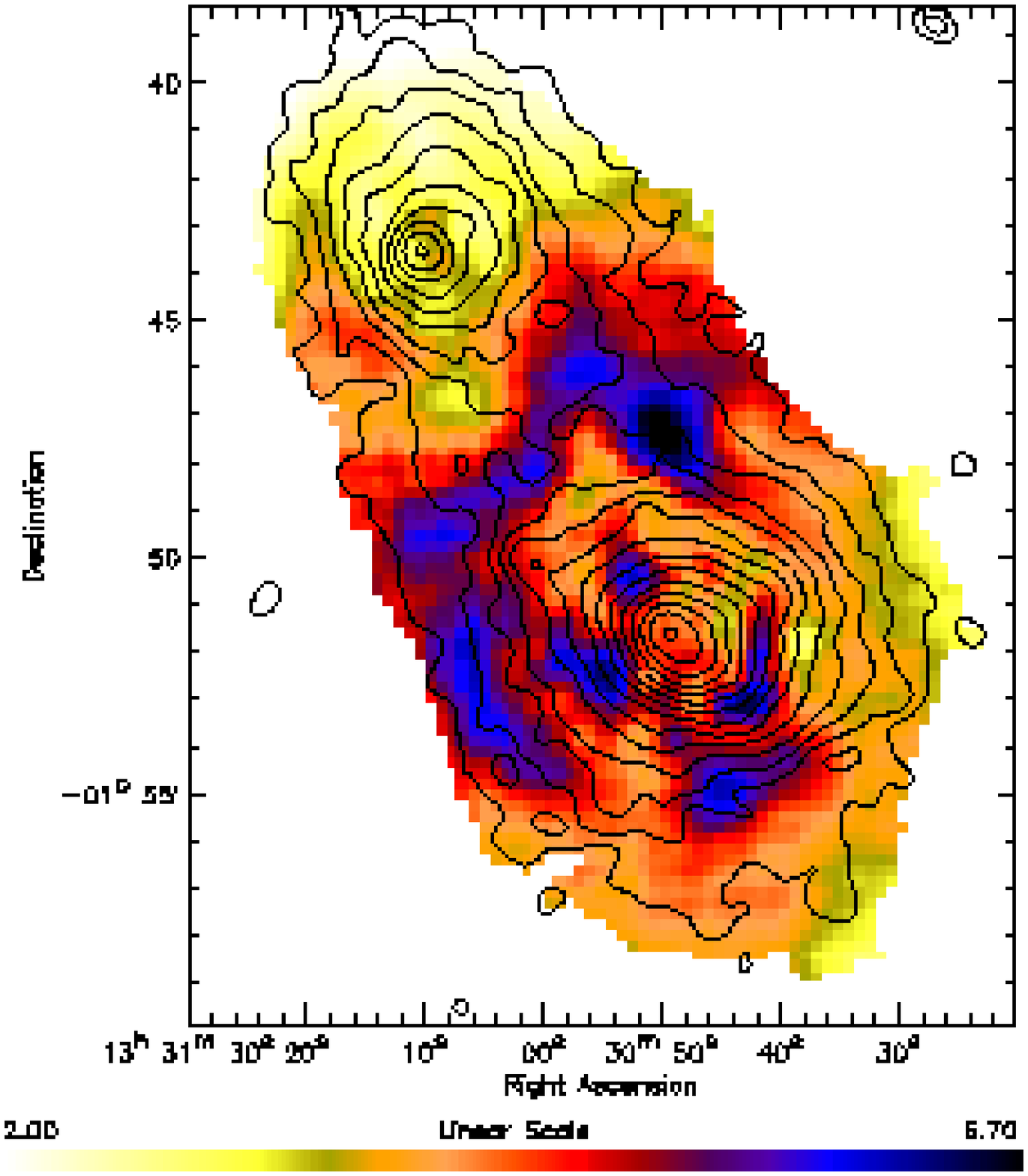}
}
\subfigure[] 
{
    \label{fig:fig3b}
    \includegraphics[scale=0.38,angle=0,keepaspectratio]{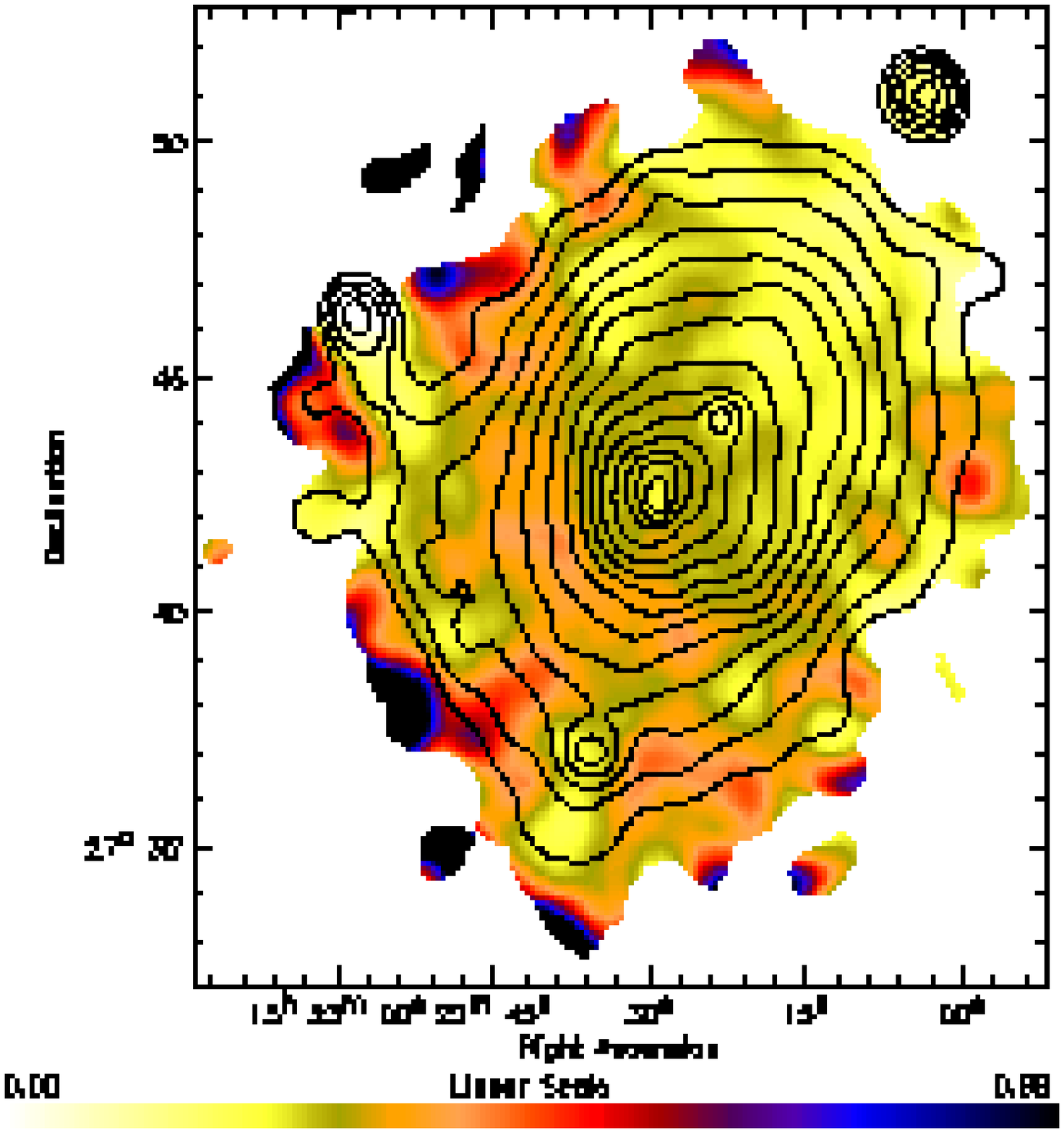}
}
\subfigure[] 
{
    \label{fig:fig3c}
    \includegraphics[scale=0.48,angle=0,keepaspectratio]{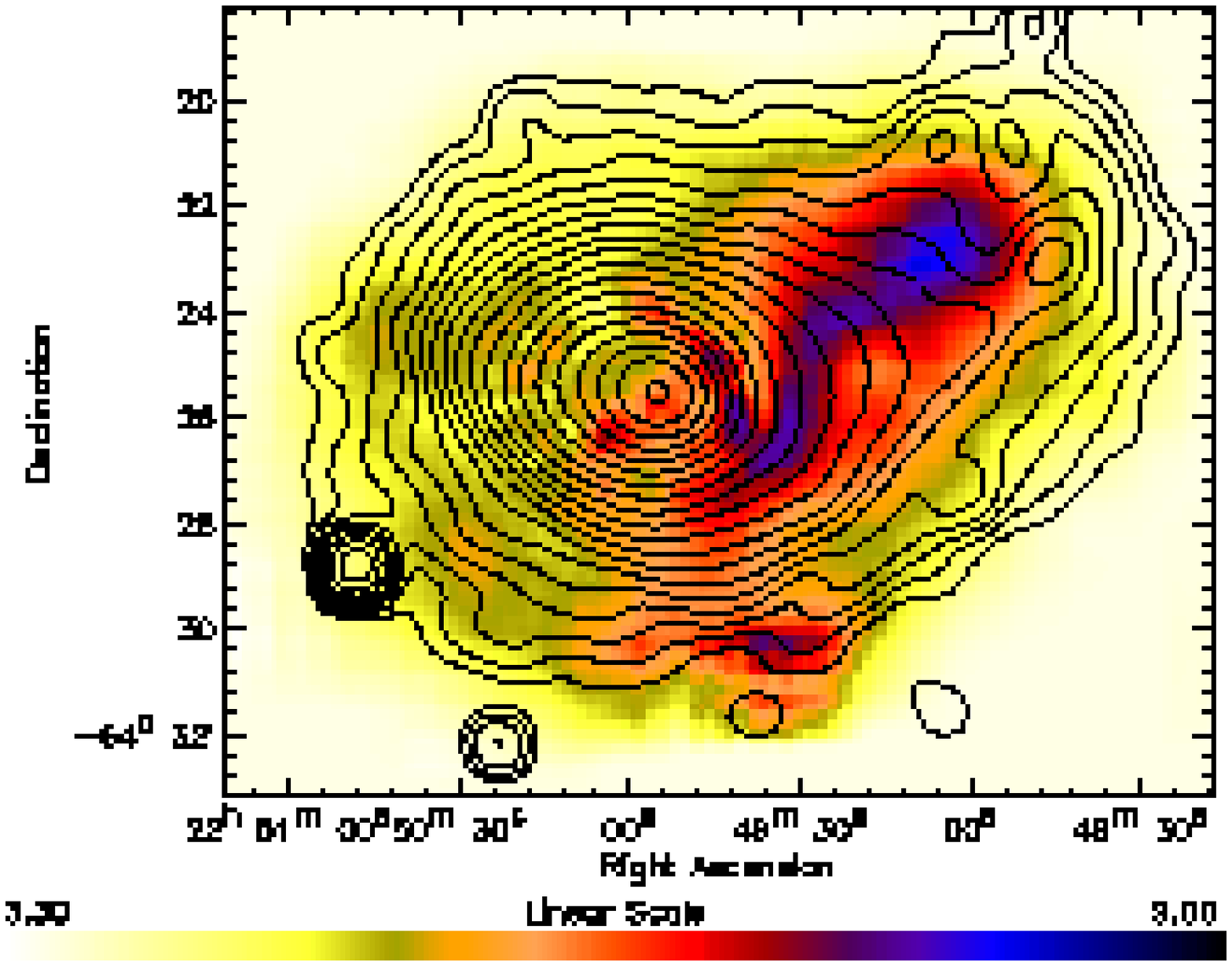}
}
\caption{{\footnotesize Temperature maps. The wavelet algorithm
    described in \citet{bourdin} was used for A1750 (a) and
    A3921(c). In the case of A2065 (b) the image is an hardness ratio map (see text for details).
   }} 
\label{fig:tmaps} 
\end{minipage}
\end{figure*}

We computed temperature maps for A1750 and A3921 by applying the
multi-scale spectro-imagery algorithm described in \citep{bourdin}. EMOS
cameras only were used for this analysis. For A2065, we obtained an
hardness ratio (HR) map once the particle contamination was modelled
and subtracted from the images. This gives us qualitative, but
significant results about temperature substructure in this cluster.
The temperature maps are shown in Figure \ref{fig:tmaps}.

All three clusters show significant temperature variations, even in
the cases where surface brightness substructures were not detected
(A1750). As is confirmed by extraction of spectra from discrete
regions \citep{belsa}, the region between A1750N and A1750C has a
temperature which 
is significantly (30\%) hotter than the global temperature of either
the two clusters. Additional temperature gradients are observed in
A1750C, indicating that this cluster was already perturbed in the past.

The temperature enhancement, coincident with the isophotal
compression, observed in the A2065 HR map points to a bow shock. We
also detect a cool core, coincident with the peak of X-ray emission. 

The orientation of the striking hot temperature bar in A3921
represents our strongest proof against the interpretation that this is
a merger before close core passage. The temperature here varies
between 7 and 8 keV, up to 60\% higher than the main cluster
temperature.

\section{Discussion}

For each cluster we summarise the results obtained in the previous sections. 

{\bf A1750}

\begin{enumerate}

\item From the morphological analysis we found: off-centre cores and
  twist of the isophotes, but lack of substructure in the region
  between the two clusters. 

\item the temperature increases weakly ($\sim30$\%), but significantly,
  in the region between the two clusters 

\item strong temperature variations are observed within the main
  cluster (A1750C) and we also measured a discontinuity in the gas
  density profile of order 20\%. 

\item we found a high entropy in the core of the two clusters (see
  \citealt{belsa})   
\end{enumerate}

The combination of these results lead us to interpret A1750 as an
ongoing merger between two clusters of similar masses, which have just
started to interact at a real distance comparable to their virial
radii. These two units will be in the compact phase (core passage)
within 1 Gyr. However the temperature and density variations within
A1750C itself are not explicable by the merger event occurring with
A1750N and are intrinsic to this cluster. The most likely
interpretation is that this is the signature of (a) previous
merger(s). A1750 is a good observational test case to be compared with
numerical simulations, since it shows all of the signatures expected
from gravitationally dominated processes. However, the fact that
effects of a previous merger are observed strongly supports the
necessity of taking into account time dependent quantities, such us
the relaxation time, in using galaxy clusters for cosmological
purposes.

\vspace{0.7cm}
{\bf A2065}

The surface brightness shows an isophotal compression towards the
south-east,  at $\sim 80 $ arcsec from the X-ray peak. The feature is accompanied by a forward shock in the same axis (NW-SE); moreover a surviving cooling core is detected. 

Despite the limited quality of the data, A2065 shows clear signatures
of being an ongoing merger in the compact phase, when the detection of
strong shocks is the most favourable. This is similar to what is
observed in numerical simulations of head-on collisions of merging
clusters. 
The fact that the core of the main cluster is cool suggests that this
is probably the remnant of a cooling core, and thus the colliding
object  was probably of smaller mass \citep[e.g.][]{gomez}. The new
XMM-Newton observations of this object should allow us to put better
constraint on these preliminary results.

{\bf A3921}

Observational merger evidence includes:
\begin{enumerate}
\item Two peaks in both the X-ray emission and the galaxy distribution
  \citep{ferrari};

\item The hot bar is oriented parallel to the line joining the
  subclusters, it is not orthogonal, as in the case of A1750; 
 
\item The central regions of the main cluster and the subcluster to
  the west are strongly perturbed; 

\item There is an off-set between the galaxy distribution of the
  smaller subcluster and the secondary X-ray peak
  \citep{belsb,ferrari}.   
\end{enumerate}

This evidence cannot be interpreted as the result of a pre-merger. We
are seeing maybe the best example of an X-ray observed off-axis
post-merger. The subcluster has come from somewhere in the SE and is
currently exiting towards the NW. The two merging units have different
masses, on a ratio of 1 to 3 (or 1 to 5; see also
\citealt{ferrari}). Comparison of these high quality X-ray results
with optical observations and numerical simulations yield an estimate
of the age of the merger of order 0.5 Gyr after core passage. Off-axis
mergers are more likely to occur than head-on ones, and they are more
efficient in mixing the gas via turbulence. If anything, A3921 needs
even deeper study, and the combination of these \xmm\ data with
upcoming  {\em Chandra} observations will give further elements to our
interpretation.

\section{Conclusions}

The \xmm\ observations of this small sample of galaxy clusters has
confirmed that the comparison of X-ray morphology and temperature is
an excellent tool to understand the dynamical status of these
objects. In the case of A1750 and A3921, spectroscopy of discrete regions (\citep{belsa,belsb} has confirmed the significance of the temperature structure found with the multi-scale wavelet approach. The significance of the temperature structure in A2065 will be investigated with the upcoming \xmm\ observation.  For the better studied cases we have confirmed the previous
interpretation of a recent merger (A1750), but have added new evidence
suggesting that at least one of the subclusters is itself a merger
remnant. Good quality X-ray data have allowed us to completely
revolutionise the interpretation of the dynamical state of A3921. 

We can organise these clusters along an evolutionary path, where A1750
represents the beginning of a merging event, which in a timescale of
order 1.5 Gyr will be in the same state as A2065, when a bow shock is
departing in the direction of motion after the two cores have
collided, and the two (or maybe more) colliding objects are not
physically separable anymore. Finally, A3921 represents the epoch when
the secondary object has already passed the close core passage phase,
it is exiting on the far side with respect to the direction of motion,
and it will be accreted by the main cluster on a larger timescale (of
order 3-5 Gyr).   

From so small a sample we cannot extract global conclusions on the
population of merging galaxy clusters. However, this work supports the
necessity of a wider investigation of the effect of the physics of
mergers on the global characteristics of galaxy clusters, especially
if these objects are to be used to derive cosmological parameters. 



\end{document}